\documentclass[12pt,draftclsnofoot,onecolumn]{IEEEtran}

\usepackage{amsfonts}
\usepackage{amssymb}
\usepackage{graphicx}
\usepackage{amsmath}
\usepackage{amsfonts, multirow, floatflt}
\usepackage{epsfig,graphics,graphicx,verbatim,slashbox}
\usepackage{latexsym,verbatim,slashbox,algorithm,algorithmic}
\usepackage{subfigure}
\usepackage{mathrsfs}
\usepackage{cite}
\usepackage{color}
\usepackage{array}
\usepackage{booktabs}
\usepackage{multicol}
\usepackage{amsthm}
\usepackage{thmtools}
\usepackage{caption}

\DeclareCaptionFormat{myformat}{#3}
\declaretheoremstyle[
headfont=\normalfont\itshape,
notefont=\normalfont, notebraces={(}{)},
bodyfont=\normalfont,
headpunct={.},
qed=\qedsymbol
]{mystyle}
\declaretheorem[style=mystyle, unnumbered]{Proof}

\begin{document}

\title{Performance of Proportional Fair Scheduling for Downlink Non-Orthogonal Multiple\\ Access Systems}

\author{Fei Liu 
and Marina Petrova
\thanks{Fei Liu is with the Institute for Networked Systems, RWTH Aachen University, 52072 Aachen, Germany (email: fei@inets.rwth-aachen.de).}
\thanks{Marina Petrova is with the School of Information and Communication Technology, KTH Royal Institute of Technology, Sweden (email: petrovam@kth.se).}
}
\maketitle
\vspace{-6ex}
\begin{center} {\bf Abstract}
\end{center}
In this paper, we present the first analytical solution for performance analysis of proportional fair scheduling (PFS) in downlink non-orthogonal multiple access (NOMA) systems. Assuming an ideal NOMA system with an arbitrary number of multiplexed users, we derive a closed-form solution of the optimal power allocation for PFS and design a low-complexity algorithm for joint power allocation and user set selection. We develop an analytical model to analyze the throughput performance of this optimal solution based on stochastic channel modeling. Our analytical performance is proved to be the upper bound for PFS and is used to estimate user data rates and overall throughput in practical NOMA systems. We conduct system-level simulations to evaluate the accuracy of our data rate estimation. The simulation results verify our analysis on the upper bound of PFS performance in NOMA and confirm that using the analytical performance for data rate estimation guarantees high accuracy. The impact of partial and imperfect channel information on the estimation performance is investigated as well.
\begin{flushleft}
\end{flushleft}
\newpage
\section{Introduction}
\label{SecIntro}
The dramatic growth of mobile data services driven by wireless Internet access and smart devices has triggered the development of 5th generation mobile networks (5G) for future wireless communications~\cite{Mag2014}. In the last generations of mobile networks, various radio access schemes were utilized as key technologies, including frequency division multiple access (FDMA), time division multiple access (TDMA), code division multiple access (CDMA), and orthogonal frequency division multiple access (OFDMA). In order to respond to the challenge of providing significantly higher system capacity in the upcoming 5G, non-orthogonal multiple access (NOMA) has been proposed and investigated as a candidate radio access technology~\cite{ISPACS2013,VTC2013}.
\par
In contrast to the previously employed orthogonal multiple access (OMA), the power-domain downlink NOMA system supports superposition coding at the transmitter side and successive interference cancellation (SIC) at the receiver side~\cite{CM2015}. Taking advantage of the diverse and time-varying user channels, power domain diversity gain is achievable with appropriate power and radio resource allocation for the multiplexed users~\cite{ISWCS2012}. Therefore, besides the resource scheduling issue, multi-user power allocation plays another key role in the performance of NOMA systems, and consequently attracts increasing research interests on it.
\par
\subsection{Related Work}
In the literature, a number of power allocation (PA) schemes have been proposed for downlink NOMA systems. They can be classified into two broad categories: fixed PA and dynamic PA. In the fixed PA schemes, the power ratios allocated to the scheduled users are preset parameters~\cite{ISWCS2012,GCWS2013}. In contrast, the dynamic PA schemes control the power ratios based on the instantaneous user channel status. One simple and well investigated dynamic PA scheme is the fractional transmit PA (FTPA)~\cite{GCWS2013,ISWCS2012,PIMRC2013YS}. The user power ratios are determined by fractional parameters calculated according to their channel qualities. Besides FTPA, several other dynamic PA schemes have been proposed with specific targets. For instance, the overall throughput in NOMA is optimized with the constraints on decoding error, data rate demand, or maximum power per user~\cite{GC2015,PIMRC2015ZQ,TSP2016}. In~\cite{SPL2015}, fairness among users has been chosen as the primary objective for PA optimization.
\par
In order to achieve a good balance between throughput and user fairness in dynamic NOMA systems, proportional fairness (PF) has been considered in many recent research papers~\cite{IEICE2014,ICC2016LF,ICC2016MJ,WCNC2016,PIMRC2015TS,TWC2016D,TWC2016L}. The optimal PA solution for proportional fair scheduling (PFS) in single-carrier NOMA (SC-NOMA) systems has been studied by a number of authors. In~\cite{ISWCS2012}, the authors designed a PA algorithm based on the iterative water-filling (IWF) method. In~\cite{IEICE2014}, a tree-searching based transmission PA (TTPA) scheme has been proposed to reduce the power searching complexity. However, both IWF and TTPA suffer from such a high complexity for the dynamic PA that they can hardly be applied in practice. The closed-form solution of the optimal PF-based PA has been derived in~\cite{ICC2016LF,PIMRC2015TS,ICC2016MJ,WCNC2016}, which reduces significantly the computational complexity of dynamic PA. In addition, the PA optimization for PFS in multi-carrier NOMA (MC-NOMA) systems were studied in~\cite{TWC2016D} and~\cite{TWC2016L}. An iterative matching algorithm was proposed in~\cite{TWC2016D} to jointly optimize subchannel and power allocation. The tractability of the optimal PA solution has been analyzed in~\cite{TWC2016L} and the authors designed a dynamic programming algorithm to obtain the near-optimal solutions.
\par
The performance of various fixed and dynamic PA schemes has mostly been evaluated by simulations. In order to further investigate and better utilize the existing PA schemes, the analytical solution of their performance is desired but has not been studied sufficiently. The outage probability and ergodic sum rate have been analyzed based on the fixed PA scheme for the single-cell networks in~\cite{SPL2014} and~\cite{TC2016} as well as for the relaying networks in~\cite{CL2015JK} and~\cite{TVT2016}. With the aim of improving the outage performance, the authors in~\cite{CL2016} have designed a dynamic PA scheme and derived the outage probabilities for the 2-user NOMA systems. However, as a widely accepted dynamic PA strategy, the performance of PFS has not been studied for NOMA analytically so far. The performance analysis of PFS in NOMA systems is important to provide guidelines for its optimization and application. In particular, the analytical results can be used for performance prediction and assisting user association, traffic load balancing, radio resource management, etc~\cite{ICC2014}. In this paper, as far as we are aware, we present the first analytical solution for PFS performance analysis in downlink NOMA systems.
\par
\subsection{Contributions}
We focus on the performance of PFS for SC-NOMA in this paper. We consider a downlink cellular network, where multiple user terminals are served dynamically with PFS.
We assume an ideal NOMA system in order to make the performance analytically tractable. The practical and ideal NOMA systems referred in this paper are defined as follows.
\begin{itemize}
\item \emph{Practical NOMA system}: The maximum number of multiplexed users is controlled by a pre-defined parameter due to the limited processing capability of SIC receivers. Normally, the limitation is 2 or 3 users in practice and the corresponding NOMA systems are referred to as 2-user and 3-user NOMA in this paper~\cite{ICC2016LF}.
\par
\item \emph{Ideal NOMA system}: We assume the SIC receiver has no limitation on the number of multiplexed users for ideal NOMA. Thus, an arbitrary number of users can be multiplexed simultaneously.
\par
\end{itemize}
\par
The main contribution of this paper is three-fold:
\begin{itemize}
  \item Based on the assumption of ideal NOMA, we derive a closed-form solution of the optimal PA for PFS. The performance of this PA solution is proved to be the upper bound for PFS in practical NOMA. Based on the derivation of the optimal PA, we design a low-complexity algorithm to jointly select the optimal multiplexed users and determine their assigned power.
      \par
  \item We develop an analytical model to analyze the throughput performance of the optimal PA solution in the ideal NOMA system. The analytical result of user data rate expectation is derived based on stochastic channel modeling. The influence of partial channel information on the analytical result is studied.
      \par
  \item Moreover, we use the analytical performance to estimate user data rates and overall throughput in the 2-user and 3-user NOMA systems. The system-level simulations are carried out to verify our analytical result of the upper bound. We also confirm that using it for data rate estimation in practical NOMA is feasible and results in very low deviations. Various influence factors on estimation accuracy, including SIC limitation, the number of users, partial and imperfect channel information, are carefully investigated.
\par
\end{itemize}
\par
The rest of the paper is organized as follows. Section II describes the system model. In Section III, we study PFS in the ideal NOMA system and derive the optimal PA solution for it. Then, its performance is analyzed in Section IV. The simulation and numerical results are presented and compared in Section V. Finally, we draw conclusions in Section VI.
\par

\section{System Model}
\label{SecModel}
We consider one base station (BS) in a downlink SC-NOMA cellular network. The BS is denoted as $b$ and the active user set associated to it is denoted as ${\bf{U}} = \left\{1,2,...,U\right\}$, where $U$ is the number of users in ${\bf{U}}$. The BS allocates radio resource and power to the users with the PFS metric in each frame.
\par
PFS considers the instantaneous user data rate along with the long-term averaged rate~\cite{VTC2000}, which is calculated as
\begin{equation}\label{eqPFR}
{R_u}\left( {t + 1} \right) = \left( {1 - \frac{1}{\tau }} \right){R_u}\left( t \right) + \frac{{{r_u}\left( t \right)}}{\tau }, \quad u\in \bf{U},
\end{equation}
\noindent where $t$ is the frame index, $\tau$ is the averaging window size, and ${r_{u}}\left(t \right)$ is the obtained data rate of user $u$ in the $t$-th frame, which depends on the system bandwidth, allocated power and user channel quality. If user $u$ is not scheduled in the $t$-th frame, ${r_{u}}\left(t \right)$ equals to 0.
\par
The scheduling metric in PFS is expressed as
\begin{equation}\label{eqw}
\omega \left( t \right) = \sum\limits_{u \in {\bf{U}}} {\frac{{{r_u}\left( t \right)}}{{{R_u}\left( t \right)}}} .
\end{equation}
\noindent In order to maximize the geometric mean of the long-term averaged user rates, $\omega \left( {t} \right)$ needs to be maximized in every frame with appropriate control of the multiplexed users and their obtained data rates~\cite{ISWCS2012}. Focusing on the PA optimization problem in one certain frame, we neglect the frame index $t$ in the following part.
\par
SIC is adopted in the NOMA system to allow superposition of multiple user signals with different transmit power levels~\cite{VTC2013}. We denote the set of users multiplexed in the considered frame as ${\bf{S}} \subseteq {\bf{U}}$. The user amount in the set is denoted as $S= \left| {\bf{S}}\right|$.
\par
Both transmission and reception use single-antenna systems. Hence, the delivered signal power from BS $b$ at a user is expressed as
\begin{equation}\label{eqrecsignal}
{P_{u,b}} = {L_{u,b}}{\left\| {{h_{u,b}}} \right\|^2}\sum\limits_{v \in {\bf{S}}} {{p_v}} ,\quad u \in {\bf{U}},
\end{equation}
\noindent where ${L_{u,b}}$ is the comprehensive channel gain, including antenna gain, path loss, and shadow fading, $h_{u,b}$ is the instantaneous Rayleigh fading gain and is modeled as a circularly symmetric complex Gaussian random variable, ${\cal{CN}}\left(0,1\right)$. Thus, its power gain, ${\left\| {h_{u,b}} \right\|^2}$, is exponentially distributed with a unit mean value. $p_v$ is the transmit power allocated to user $v \in {\bf{S}}$, which can be further expressed as
\begin{equation}\label{eqpowerratio}
{p_u} = {\lambda _u}{p_T},\quad u \in {{\bf{S}}},
\end{equation}
\noindent where $p_T$ is the total transmit power, and ${\lambda _u}$ is the power ratio assigned to user $u$ and satisfies
\begin{equation}\label{eqratiocondition}
\sum\limits_{u \in {\bf{S}}} {{\lambda _u}}  = 1, \quad {\rm{and}} \quad {\lambda _u} \in \left( {0,1} \right].
\end{equation}
\par
We denote the channel quality indicator (CQI) of user $u$ as ${\Phi_u}$. It is defined as the signal-to-interference-plus-noise ratio (SINR) while the full transmit power of the BS, i.e., $p_T$, is allocated to the user. Hence, ${\Phi _u}$ is calculated as
\begin{equation}\label{eqSINR}
{\Phi _u} = {{{\hat P_{u,b}}} \over {\sum\limits_{i \in {{\bf{I}}_u}} {{P_{u,i}} + {\sigma _u}} }},
\end{equation}
where
\begin{equation}\label{eqfullpower}
{{\hat P}_{u,b}} = {L_{u,b}}{\left\| {{h_{u,b}}} \right\|^2}{p_T},
\end{equation}
\noindent ${\bf{I}}_u$ denotes the inter-cell interferer set of user $u$, $P_{u,i}$ denotes the received inter-cell interference power from BS $i$, and ${\sigma _u}$ is the additive noise received by user $u$. The inter-cell interference cancellation is not considered. Thus, the received $P_{u,i}$ has an adverse impact on CQI.
\par
\begin{figure}[t]
\centering
{\includegraphics[width=3.1in]{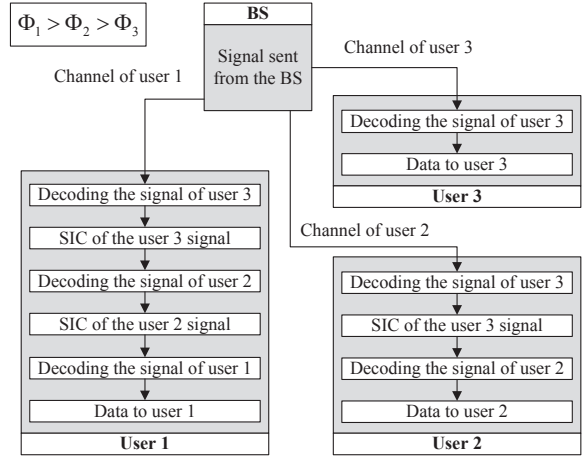}}
\caption{Illustration of SIC for 3-user NOMA transmission.}
\label{Fig3users}
\end{figure}
In downlink NOMA systems, SIC is carried out in user receivers for decoding. For ease of derivation, we assume that the multiplexed users in the frame are sorted in the descending order in terms of their instantaneous CQIs, which is defined as a scheduled user sequence,
\begin{equation}
\mathbb{S} = \Theta \left( {{\bf{S}}} \right) =  \left\langle {c_1,c_2, \ldots ,c_{S}} \right\rangle ,
\end{equation}
\noindent where
\begin{equation}
{\Phi _{c_n}} \ge {\Phi _{c_{n + 1}}},\quad n = 1, \ldots ,{S} - 1,
\end{equation}
\noindent and $\Theta \left(\cdot  \right)$ is the operator for descending sorting of the user indices in terms of their CQIs. 
The $n$-th user in $\mathbb{S} $ decodes and cancels successively the interference signals of user $c_{n + 1} \sim c_{{S}}$ in reverse order by using SIC~\cite{VTC2013}.
\par
In Fig.~\ref{Fig3users}, we present an example of the reception and decoding process in the 3-user NOMA system. User 1 has the highest CQI thus needs to firstly decode and cancel the interference signal of the other two users. Then, it decodes its own signal. User 2 carries out SIC to eliminate the interference signal of user 3 and regards the signal of user 1 as noise. No SIC is necessary at user 3 for decoding its signal.
\par
The post-processing SINR of a scheduled user $c_{n}$ after SIC is calculated as
\begin{equation}\label{eqpostSINR}
{\gamma _{{c_n}}} = \left\{ {\begin{array}{*{20}{c}}
   {{\Phi _{{c_n}}}{\lambda _{{c_n}}},} & {n = 1,}  \\
   \displaystyle{\frac{{{\Phi _{{c_n}}}{\lambda _{{c_n}}}}}{{{\Phi _{{c_n}}}\sum\limits_{m = 1}^{n - 1} {{\lambda _{{c_m}}}}  + 1}},} & {n = 2, \ldots ,S.}  \\
\end{array}} \right.
\end{equation}
\noindent For implementation of SIC, a user with better channel quality has to be informed of the modulation and coding schemes and the power ratios allocated to the users that have lower CQIs. If $S$ is large, the signalling cost and processing complexity at user terminals become very high. Therefore, $S$ is not larger than a predefined limitation $S_{max}$ in practice. Particularly, we have $S_{max}=1$ in OMA systems.
\par
\section{Optimal PFS in Ideal NOMA Systems}
In this section, we derive a closed-form solution of the optimal PA in the ideal NOMA system, where $S_{max} = \infty$. Without the limitation on the number of multiplexed users $S$, this solution results in an upper bound of PFS performance. In order to select the corresponding multiplexed users, we design a low-complexity algorithm in the next step according to our derivation.
\par
\subsection{Derivation of the Optimal PA Solution}
We define the cumulative power ratio (CPR) allocated to the first $n$ users in the sorted user sequence $\mathbb{S}$ as
\begin{equation}
{\alpha _n} = \sum\limits_{m = 1}^n {{\lambda _{c_m}}} ,\quad n = 1, \ldots ,{S}.
\end{equation}
\noindent Without loss of generality, we define ${\alpha _0} = 0$ for ease of expression in the following derivation. According to \eqref{eqratiocondition}, we have ${\alpha _{{S}}} = 1$ and
\begin{equation}\label{eqrelationshipalpha}
{\alpha _{n - 1}} < {\alpha _n},\quad n = 1, \ldots ,{S}.
\end{equation}
\par
With the above relationship, we define a sequence of CPRs in ascending order as
\begin{equation}
\mathbb{A} = \left\langle {{\alpha _0},{\alpha _1} \ldots ,{\alpha _S}} \right\rangle .
\end{equation}
\par
Then, the PFS metric in \eqref{eqw} can be rewritten as
\begin{equation}\label{eqw2}
\omega \left( \mathbb{S}, \mathbb{A}\right) = \sum\limits_{n = 1}^S {\frac{{{r_{{c_n}}}\left( \mathbb{A} \right)}}{{{R_{{c_n}}}}}} ,
\end{equation}
\noindent where ${{r_{c_n}}\left( {\mathbb{A}} \right)}$ is the instantaneous obtainable data rate of user $c_n$ and is calculated by the Shannon capacity. Hence, we have
\begin{equation}\label{eqrshannon}
{r_{{c_n}}}\left( {\mathbb{A}} \right) = {B}{\log _2}\left( {\frac{{1 + {\alpha _n}{\Phi _{{c_n}}}}}{{1 + {\alpha _{n - 1}}{\Phi _{{c_n}}}}}} \right),\quad n = 1, \ldots ,{S},
\end{equation}
\noindent where $B$ is the system bandwidth.
\par
Combining \eqref{eqw2} and \eqref{eqrshannon}, the PFS metric is calculated as
\begin{align}\label{eqPFfactor}
 \omega \left( {{\mathbb{S}},{\mathbb{A}}} \right) =& {B}\sum\limits_{n = 1}^S {\left[ {{{\log }_2}\left( {1 + {\alpha _n}{\Phi _{{c_n}}}} \right) - {{\log }_2}\left( {1 + {\alpha _{n - 1}}{\Phi _{{c_n}}}} \right)} \right]} R_{{c_n}}^{ - 1} \notag\\
  {\mathop  = \limits^{\eqref{eqrelationshipalpha}} }& \frac{{{B}}}{{\ln 2}}\sum\limits_{n = 1}^S {\int_{{\alpha _{n - 1}}}^{{\alpha _n}} {\frac{{{\Phi _{{c_n}}}}}{{{R_{{c_n}}}\left( {1 + x {\Phi _{{c_n}}}} \right)}}d x } }.
\end{align}
We denote the smooth coefficient function (CF) in the integral part of \eqref{eqPFfactor} as
\begin{equation}\label{eqfunctionf}
{\pi_u}\left( x \right) = \frac{{{\Phi _u}}}{{{R_u}\left( {1 + x{\Phi _u}} \right)}},\quad x\in \left( {0,1} \right) .
\end{equation}
Substituting \eqref{eqfunctionf} into \eqref{eqPFfactor}, we deduce the maximum PFS metric as follows,
\begin{align}\label{eqwint}
 \omega \left( {{\mathbb{S}},{\mathbb{A}}} \right) =& \frac{{{B}}}{{\ln 2}}\sum\limits_{n = 1}^S {\int_{{\alpha _{n - 1}}}^{{\alpha _n}} {{\pi_{{c_n}}}\left( x  \right)d x } } \notag \\
  \le& \frac{{{B}}}{{\ln 2}}\sum\limits_{n = 1}^S {\int_{{\alpha _{n - 1}}}^{{\alpha _n}} {\mathop {\max }\limits_{m = 1}^S {\pi_{{c_m}}}\left( x  \right)d x } }  \\
  =& \frac{{{B}}}{{\ln 2}}\int_0^1 {\mathop {\max }\limits_{m = 1}^S {\pi_{{c_m}}}\left(  x  \right)d x }  \notag\\
  \le& \frac{{{B}}}{{\ln 2}}\int_0^1 {\mathop {\max }\limits_{u \in {\bf{U}} } {\pi_u}\left( x  \right)d x }  \notag
\end{align}
\noindent Thus, the maximum PFS metric is calculated as
\begin{equation}\label{eqmaxPFfactor}
{\omega^*}=\frac{{{B}}}{{\ln 2}}\int_0^1 {\mathop {\max }\limits_{u \in {{\bf{U}}}} {\pi_u}\left( x \right)d x }.
\end{equation}
\par
To express the selected user set for obtaining the maximum PFS metric, we define a user index selection function as
\begin{equation}\label{eqselecteduser}
l\left( x \right) = \mathop {\arg \max }\limits_{u \in {{\bf{U}}}} {\pi_u}\left( x \right).
\end{equation}
\noindent We denote the optimal scheduled user set as
\begin{equation}\label{eqselecteduserset}
{{\bf{ S}}^*} = \left\{ {l\left( x \right)\left| x \in \left( {0,1} \right) \right.} \right\}.
\end{equation}
\noindent The corresponding optimal scheduled user sequence is
\begin{equation}\label{eqoptimalusersequence}
 \mathbb{S}^* = \Theta \left( {{{{\bf{ S}}}^*}} \right) = \langle {{ c^*_1},{ c^*_2}, \ldots ,{ c^*_{ S^*}}} \rangle,\quad {S^*} = \left| {\bf{S}}^* \right|.
\end{equation}
\par
We denote the optimal CPR sequence for maximizing the PFS metric as
\begin{equation}
{\mathbb{A}^*} = \left\langle {\alpha _0^{\rm{*}},\alpha _1^{\rm{*}}, \ldots ,\alpha _{{S^*}}^{\rm{*}}} \right\rangle,
\end{equation}
\noindent in which $\alpha^*_0$ is fixed to 0, and $\alpha^*_{S^*}$ is naturally 1. In order to solve ${\mathbb{A}^*}$, we utilize the theorem and corollaries presented as follows.
\par
Firstly, we consider the condition that ${\Phi _u} \ne {\Phi _v},\; \forall u,v \in {{\bf{U}}}$.
\par
\textbf{Theorem 1.}
\emph{For two users who have ${\Phi _u} > {\Phi _v},\; u,v \in {{\bf{U}}}$,}
\par
\begin{description}
  \item[\emph{Case 1)}] \quad \emph{if they meet the condition:}
  \begin{equation}\label{eqconditionalpha}
    0 < \frac{{{\Phi _v}}}{{{\Phi _u}}} < \frac{{{R_v}}}{{{R_u}}} < \frac{{1 + \Phi _u^{ - 1}}}{{1 + \Phi _v^{ - 1}}} < 1,
  \end{equation}
  \emph{then it holds that}
  \begin{equation}
\begin{array}{*{20}{c}}
   {{\pi _u}\left( x \right) = {\pi _v}\left( x \right),} & {x = {\theta _{u,v}},}  \notag \\
   {{\pi _u}\left( x \right) > {\pi _v}\left( x \right),} & {x \in \left( {0,{\theta _{u,v}}} \right),}  \notag\\
   {{\pi _u}\left( x \right) < {\pi _v}\left( x \right),} & {x \in  \left( {{\theta _{u,v}},1} \right),}  \notag\\
\end{array}
  \end{equation}
  \emph{where}
  \begin{equation}\label{eqalpha}
{\theta_{u,v}} = {\theta_{v,u}} = \frac{{{R_u}\Phi _u^{ - 1} - {R_v}\Phi _v^{ - 1}}}{{{R_v} - {R_u}}};
\end{equation}
  \item[\emph{Case 2)}] \quad \emph{if they meet the condition:}
    \begin{equation}
    \frac{{{R_v}}}{{{R_u}}} \le \frac{{{\Phi _v}}}{{{\Phi _u}}},
  \end{equation}
  \emph{then it holds that}
   \begin{equation}
   {\pi_u}\left( x \right) < {\pi_v}\left( x \right),\quad x \in \left( {0,1} \right); \notag
   \end{equation}
  \item[\emph{Case 3)}] \quad \emph{if they meet the condition:}
  \begin{equation}\label{eqconditiona3}
    \frac{{{R_v}}}{{{R_u}}} \ge \frac{{1 + \Phi _u^{ - 1}}}{{1 + \Phi _v^{ - 1}}},
  \end{equation}
   \noindent\emph{ then it holds that}
   \begin{equation}
   {\pi_u}\left( x \right) > {\pi_v}\left( x \right),\quad x \in \left( {0,1} \right). \notag
   \end{equation}
\end{description}
\par
The proof of Theorem 1 is given in Appendix~A. Based on this theorem, we deduce two corollaries as follows.
\par
\textbf{Corollary 1.}
\emph{In the optimal scheduled user sequence $\mathbb{S}^*$, it holds that}
\begin{align}\label{eqrelationships}
   {{\pi _{c_n^*}}\left( x \right) > {\pi _{c_{n + 1}^*}}\left( x \right),} & \quad{x \in \left( {0,{\theta _{c_n^*,c_{n + 1}^*}}} \right),} \notag \\
   {{\pi _{c_n^*}}\left( x \right) < {\pi _{c_{n + 1}^*}}\left( x \right),}  &\quad{x \in \left( {{\theta _{c_n^*,c_{n + 1}^*}},1} \right),}  \notag\\
   n =1,2, \ldots ,&{S^*}-1. \notag
\end{align}
\par
\begin{Proof}
According to~\eqref{eqselecteduser}, \eqref{eqselecteduserset} and~\eqref{eqoptimalusersequence}, it is obvious that
\begin{equation}\label{eqC1proof}
\begin{aligned}
&\forall {c_n^*} \in \mathbb{S}^*,\quad \exists \, {y} \in \left( {0,1} \right),  \\
{\rm{s.t.,}}\quad &{\pi _{c_n^*}}\left( y \right) > {\pi _{c_m^*}}\left( y \right),\quad m\ne n. 
\end{aligned}
\end{equation}
\noindent Due to this fact, every pair of scheduled users in $\mathbb{S}^*$ must meet condition \eqref{eqconditionalpha} in Case 1 of Theorem 1. Otherwise, one of them must has lower CF than the other as shown in Case 2 or 3 of Theorem 1 thus is not selected by \eqref{eqselecteduserset}. Therefore, according to Case 1 of Theorem 1, we have the CF relationships between two adjacent users in $ \mathbb{S}^*$ as shown in Corollary 1.
\end{Proof}
\par
\textbf{Corollary 2.}
\emph{In the optimal scheduled user sequence $\mathbb{S}^*$, it holds that}
\begin{equation}\label{eqCorollary2}
{\theta _{c_{n - 1}^*,c_n^*}} < {\theta _{c_n^*,c_{n + 1}^*}},\quad n = 2, \ldots ,{S^*} - 1.
\end{equation}
\begin{Proof}
According to Corollary 1, we have
\begin{align}
{{\pi _{c_n^*}}\left( x \right) < {\pi _{c_{n - 1}^*}}\left( x \right),} & \quad{x \in \left( {0,{\theta _{c_{n-1}^*,c_{n}^*}}} \right)},\label{eqrelationshipsC2proof1}\\
{{\pi _{c_n^*}}\left( x \right) < {\pi _{c_{n + 1}^*}}\left( x \right),} & \quad{x \in \left( {{\theta _{c_n^*,c_{n + 1}^*}},1} \right)},\label{eqrelationshipsC2proof2}\\
n =2, \ldots ,&{S^*}-1. \notag
\end{align}
Assuming that the $n$-th user in $\mathbb{S}^*$ leads to
\begin{equation}\label{C2proof}
{\theta _{c_{n - 1}^*,c_n^*}} \ge {\theta _{c_n^*,c_{n + 1}^*}},
\end{equation}
due to~\eqref{eqrelationshipsC2proof1} and~\eqref{eqrelationshipsC2proof2}, it is clear that
\begin{align}
&\nexists \, {y} \in \left( {0,1} \right), \notag \\
{\rm{s.t.,}}\quad &{\pi _{c_n^*}}\left( y \right) > {\pi _{c_m^*}}\left( y \right),\quad m \ne n. \notag 
\end{align}
\noindent This conflicts with~\eqref{eqC1proof}. Therefore, user $c_n^*$ cannot be selected by~\eqref{eqselecteduserset} as a scheduled user in $\mathbb{S}^*$. Thus, assumption~\eqref{C2proof} is false and~\eqref{eqCorollary2} holds.
\end{Proof}
\par
By Corollary 1 and 2, formula~\eqref{eqselecteduser} can be expanded into
\begin{equation}\label{eqselectedusers2}
l\left( x \right) = \left\{ {\begin{array}{*{20}{c}}
   {c_1^*,} & {x \in \left( {0,{\theta _{c_1^*,c_2^*}}} \right),}  \\
   {c_n^*,} & \begin{array}{l}
 x \in \left( {{\theta _{c_{n - 1}^*,c_n^*}},{\theta _{c_n^*,c_{n + 1}^*}}} \right) \\
 {\rm{and}} \quad n = 2, \ldots ,{S^*} - 1, \\
 \end{array}  \\
   {c_{{S^*}}^*,} & {x \in \left( {{\theta _{c_{{S^*} - 1}^*,c_{{S^*}}^*}},1} \right)}  .\\
\end{array}} \right.
\end{equation}
\par
Thus, the maximum PFS metric is the integral of the multiple CFs that are maximum along $S^*$ adjacent regions in $x\in\left(0,1\right)$. We can calculate it as follows,
\begin{align}\label{eqoptimalweight}
{\omega^*} &= \frac{{{B}}}{{\ln 2}}\int_0^{{\theta _{c_1^*,c_2^*}}} {{\pi _{c_1^*}}\left( x \right)} dx \notag\\
  &+ \frac{{{B}}}{{\ln 2}}\sum\limits_{n = 2}^{{S^*} - 1} {\int_{{\theta _{c_{n - 1}^*,c_n^*}}}^{{\theta _{c_n^*,c_{n + 1}^*}}} {{\pi _{c_n^*}}\left( x \right)} dx}\notag \\
  &+ \frac{{{B}}}{{\ln 2}}\int_{{\theta _{c_{{S^*} - 1}^*,c_{{S^*}}^*}}}^1 {{\pi _{c_{{S^*}}^*}}\left( x \right)} dx.
\end{align}
\par
Comparing~\eqref{eqoptimalweight} to~\eqref{eqwint}, it is a clear fact that the optimal CPRs are
\begin{equation}\label{eqoptimalalpha}
\alpha _n^* = \theta_{ {c_n^*,c_{n + 1}^*} },\quad n = 1, \ldots ,{S^*} - 1.
\end{equation}
\par
So far, we have solved the optimal user sequence $\mathbb{S}^*$ and CPR sequence $\mathbb{A}^*$ for the maximum PFS metric, i.e.,
\begin{equation}\label{eqoptimalwww}
{\omega^*} = \omega \left( \mathbb{S}^*,\mathbb{A}^* \right).
\end{equation}
\par
We considered the condition that ${\Phi _u} \ne {\Phi _v}$ in the analysis above. Although the possibility of ${\Phi _u} = {\Phi _v}$ is very low in realistic systems, we consider this special case in the following part for the sake of analysis completeness.
\par
\textbf{Theorem 2.} \emph{For two users who have ${\Phi _u}= {\Phi _v},\; u,v \in {{\bf{U}}}$,}
\begin{enumerate}
  \item \emph{if ${R_v} < {R_u}$, then ${\pi_u}\left( x \right) < {\pi_v}\left( x \right)$;}
      \item \emph{if ${R_v} > {R_u}$, then ${\pi_u}\left( x \right) > {\pi_v}\left( x \right)$;}
   \item \emph{if ${R_v} = {R_u}$, then ${\pi_u}\left( x \right) = {\pi_v}\left( x \right)$.}
\end{enumerate}
\par
\begin{proof}
If ${\Phi _u}= {\Phi _v}$, we have
\begin{equation}
\frac{{{\pi _u}\left( x \right)}}{{{\pi _v}\left( x \right)}} = \frac{{{R_v}}}{{{R_u}}}.
\end{equation}
\noindent Thus, the three cases in Theorem 2 obviously hold.
\end{proof}
\par
According to Theorem 2, when two users have the same level of CQI, only one of them will be scheduled. Coincidentally, when they also have the identical averaged data rate, we can randomly select one of them for scheduling. This choice has no influence on maximizing the PFS metric in~\eqref{eqwint}. Thus, Theorem 2 indicates that the scheduled users in $\mathbb{S}^*$ must have mutually different CQIs.
\par
The optimal solution presented in~\eqref{eqoptimalwww} is based on the assumption that $S_{max} = \infty$. However, $S_{max}$ is normally a small number in practical NOMA systems as we discussed in Section II. Thus, the number of scheduled users, $S$, is limited by $S_{max}$, and it is possible that $S^*>S_{max} \ge S$. If so, the maximum PFS metric in~\eqref{eqoptimalwww} cannot be guaranteed with a small $S_{max}$. Therefore, the performance of the ideal NOMA system with $S_{max} = \infty$ is the upper bound for PFS in practice.
\par
\subsection{Algorithm for Optimal PFS}
Based on the analysis above, we design now a low-complexity algorithm to calculate $\mathbb{S}^*$ and $\mathbb{A}^*$ jointly in Algorithm~1. The computational complexity of Algorithm 1 is no more than $\left(U+1\right)U/2$ for user set selection, thus is much lower than $\left(2^U-1\right)$ in the full user set comparison method~\cite{ISWCS2012,IEICE2014}. Moreover, this algorithm has the advantage that it is unnecessary to sort the users in terms of their CQI before its execution. Therefore, the computational complexity can be further reduced.
\par
\begin{figure}[t]
\centering
{\includegraphics[width=3.1in]{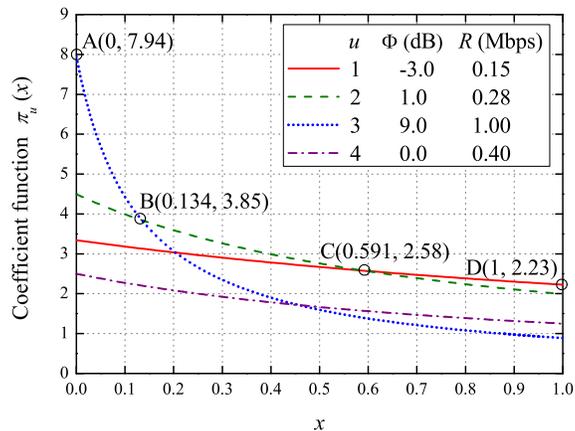}}
\caption{A 4-user example of the CF curves.}
\label{Schematic}
\end{figure}

We present a 4-user example of their CF curves in Fig.~\ref{Schematic}. By using Algorithm~1, we obtain the optimal scheduled user sequence, $\mathbb{S}^* = \langle 3,\,2,\,1\rangle$, and corresponding CPRs, $\mathbb{A}^*=\langle 0,\,0.134,\,0.591,\,1\rangle$. The maximum PFS metric is the integral along the segments A-B-C-D. User 2 and 4 meet condition~\eqref{eqconditiona3} in Theorem 1. Therefore, user 4 has a lower CF than user 2 within $x\in\left(0,1\right)$ and must not be selected for the optimal solution.
\par
\begin{algorithm}[H]
\caption{{\bf{Algorithm 1}} Optimal PFS in Ideal NOMA}
\label{alg:A}
\begin{algorithmic}[1]
\STATE $c_1 = \mathop {\arg \max }\limits_{u \in {\bf{U}}} \pi_u\left( 0 \right),$
\STATE $ \mathbb{S}_1 = \langle {c_1}\rangle,$ $\quad \mathbb{A}_1 = \langle 0\rangle,$
\STATE ${{\bf{V}}_1} = \left\{ \left. u \right|\theta_{{c_1},u}  \in \left( {0,1} \right),u \in \left( {\bf{U}}/c_1\right)\right\} ,$
\STATE $n=1.$
\WHILE {${{\bf{V}}_n} \ne \varnothing$}
\STATE ${c_{n + 1}} = \mathop {\arg \min }\limits_{u \in {{\bf{V}}_n}} \theta_{{c_n},u},$
\STATE $\mathbb{S}_{n+1} = \langle\mathbb{S}_n,c_{n + 1}\rangle,$ $\quad \mathbb{A}_{n+1} = \langle\mathbb{A}_n,\theta_{c_n,c_{n + 1}}\rangle,$
\STATE ${{\bf{V}}_{n + 1}} = \left\{ \left. u \right|\theta_{{c_{n+1},u}}  \in \left( {0,1} \right),u \in \left( {{\bf{V}}_n}/c_{n+1}\right)\right\} ,$
\STATE $n= n+1.$
\ENDWHILE
\STATE $\mathbb{S}^* = \mathbb{S}_{n},$ $\quad \mathbb{A}^* = \langle\mathbb{A}_{n},1\rangle.$\\
\COMMENT {Note: If there are multiple maximums in line 1 or multiple minimums in line 6, the algorithm chooses the user with the smallest CQI, $\Phi_u$, due to Corollary 1. If multiple users have the same smallest CQI, it randomly chooses one of them due to Theorem 2.}
\end{algorithmic}
\end{algorithm}
\section{Performance Analysis of PFS}
In this part, we analyze the throughput performance of the optimal PA solution developed for ideal NOMA in the last section. The analytical solution of user data rate expectation is derived based on stochastic SINR modeling.
\par
\subsection{Expectation of User Data Rate}
Under the fluctuant radio channels, the CQI of a user is a random variable. We denote the cumulative distribution function (CDF) and probability distribution function (PDF) of CQI as
\begin{align}\label{eqCDFPDFCQI}
{F_{{\Phi _u}}}\left( \phi  \right) =& {\rm{P}}\left\{ {{\Phi _u} < \phi } \right\},\\
{f_{{\Phi _u}}}\left( \phi  \right) =& \frac{{\partial {F_{{\Phi _u}}}\left( \phi  \right)}}{{\partial \phi }}, \quad \phi>0.
\end{align}
\par
Thus, the CF defined in~\eqref{eqfunctionf} is a random variable depending on $\Phi _u$, $R _u$ and $x$. We derive its conditional CDF as follows,
\begin{align}\label{eqCDFCF}
 {F_{{\pi _u}}}\left( {g\left| x \right.} \right) =& {\rm{P}}\left\{ {{\pi _u}\left( x \right) < g} \right\} \\
  =& {\rm{P}}\left\{ {\frac{{{\Phi _u}}}{{\left( {{\Phi _u}x + 1} \right){R_u}}} < g} \right\} \notag\\
  =& {\rm{P}}\left\{ {{\Phi _u} < \frac{{g{R_u}}}{{1 - g{R_u}x}}} \right\} \notag\\
  =& {F_{{\Phi _u}}}\left( {\frac{{g{R_u}}}{{1 - g{R_u}x}}} \right),\notag\\
   g \in &\left( {0,\frac{1}{{x{R_u}}}} \right),\quad x\in\left(0,1\right). \notag
\end{align}
Consequently, the conditional PDF of the CQI is derived as
\begin{equation}\label{eqPDFCF}
{f_{{\pi_u}}}\left( {g \left| x \right.} \right) = {f_{{\Phi _u}}}\left( {\frac{{g{R_u}}}{{1 - g{R_u}x}}} \right)\frac{{{R_u}}}{{{{\left( {1 - g{R_u}x} \right)}^2}}}.
\end{equation}
\par
We denote the expectation of user data rate as
\begin{equation}\label{eqmeanrate}
{{\overline r}_u} = \mathbb{E}\left[ {{r_u}} \right],\quad u\in{\bf{U}}.
\end{equation}
Thus, the overall throughput is expressed as
\begin{equation}\label{}
{\overline r_\Sigma } = \sum\limits_{u \in {\bf{U}}} {{{\overline r}_u}}.
\end{equation}
\par
When $\tau\gg 1$, we have the approximation that
\begin{equation}\label{eqapproximationrate}
{{\overline r}_u} = \mathbb{E}\left[ {{R_u}} \right] \approx {R_u}.
\end{equation}
\noindent This approximation is proved in Appendix~B. Thus, the expectation of the PFS metric approximately equals to 1, i.e.,
\begin{equation}\label{eqwapproxi}
{{\overline \omega }_u} = \mathbb{E}\left[ {\frac{{{r_u}}}{{{R_u}}}} \right] \approx 1.
\end{equation}
\par
On the other hand, we can calculate ${{\overline \omega }_u}$ with~\eqref{eqCDFCF} and~\eqref{eqPDFCF} as follows,
\begin{align}\label{eqwapproxi2}
 {{\overline \omega }_u} = & \int_0^1 {\int_0^{\frac{1}{{x{R_u}}}} {{f_{{\pi_u}}}\left( {g \left| x \right.} \right)\prod\limits_{v \in \left( {{\bf{U}}/u} \right)} {{F_{{\pi_v}}}\left( {g \left| x \right.} \right)} } \,g\,dg\,dx}  \notag \\
 \mathop  \approx \limits^{\eqref{eqapproximationrate}} &\int_0^1 {\int_0^{\frac{1}{{x{{\overline r}_u}}}} {{f_{{\Phi _u}}}\left( {\frac{{g{{\overline r}_u}}}{{1 - g{{\overline r}_u}x}}} \right)\frac{{g{{\overline r}_u}}}{{{{\left( {1 - g{{\overline r}_u}x} \right)}^2}}}} }  \notag\\
  & \times \prod\limits_{v \in \left( {{\bf{U}}/u} \right)} {{F_{{\Phi _v}}}\left( {\frac{{g{{\overline r}_v}}}{{1 - g{{\overline r}_v}x}}} \right)\,dg\,dx} \notag  \\
 \mathop  \approx \limits^{\eqref{eqwapproxi}}& 1,
\end{align}
\noindent which is the mean value of ${{\omega }_u}$ under the condition that user $u$ is selected by~\eqref{eqselecteduserset}. Assuming ergodicity of the radio channels, the expectation of user data rate ${\overline r}_u$ can be obtained by solving~\eqref{eqwapproxi2}.
\par
\subsection{Stochastic SINR Model}\label{SINRmodel}
In order to solve~\eqref{eqwapproxi2}, we use the stochastic SINR models in~\cite{SECON2015} to formulate ${F_{{\Phi _u}}}\left( \phi  \right)$ and ${f_{{\Phi _u}}}\left( \phi  \right)$ for multi-cell scenarios, which are presented as follows,
\begin{equation}\label{eqSINRCDFMIA}
{F_{{\Phi _u}}}\left( \phi  \right) = 1 - \exp \left( { - \frac{{{\sigma _u}}}{{{{\hat p}_{u,b}}}}\phi } \right)\prod\limits_{i \in {{\bf{I}}_u}} {{{\left( {\frac{{{p_{u,i}}}}{{{{\hat p}_{u,b}}}}\phi  + 1} \right)}^{ - 1}}},
\end{equation}
\begin{equation}\label{eqSINRPDFMIA}
{f_{{\Phi _u}}}\left( \phi  \right) = \left( {\frac{{{\sigma _u}}}{{{{\hat p}_{u,b}}}} + \sum\limits_{i \in {{\bf{I}}_u}} {\frac{{{p_{u,i}}}}{{{p_{u,i}}\phi  + {{\hat p}_{u,b}}}}} } \right)\left[ {1 - {F_{{\Phi _u}}}\left( \phi  \right)} \right],
\end{equation}
\noindent where
\begin{equation}\label{eqRSRP}
{{\hat p}_{u,b}} = \mathbb{E}\left[ {{{\hat P}_{u,b}}} \right] = {L_{u,b}}{p_T},
\end{equation}
\begin{equation}\label{eqIRSRP}
{p_{u,i}} = \mathbb{E}\left[ {{P_{u,i}}} \right] = {L_{u,i}}{p_T},\quad i \in {{\bf{I}}_u}.
\end{equation}
\par
In cellular networks, a user terminal can measure the reference signal received powers (RSRPs) of its associated BS and relatively strong inter-cell interferers nearby~\cite{3GPP36214}. It reports periodically the RSRP of BS $b$ and at most $I_{max}$ largest inter-cell interfering RSRPs (IRSRPs) in ${{{{\bf{I}}}_u}}$. The BS set including the $I_{max}$ reported interferers is denoted as ${{{{\bf{I}}}'_u}}\subseteq{{{{\bf{I}}}_u}}$. Hence, the reported RSRPs can be used to estimate ${{\hat p}_{u,b}}$ and ${p_{u,i}},\;i\in{\bf{I'}}_u$. Particularly, in the single-cell scenario, we have ${\bf{I}}'_u={\bf{I}}_u=\varnothing$.
\par
We denote the sum power of the unreported weaker inter-cell interference signals and additive noise as
\begin{equation}\label{eqasnoise}
{{p}'_{u,\sigma }} = \sum\limits_{i \in \left( {{{\bf{I}}_u} - {{{\bf{I}}}'_u}} \right)} {{p_{u,i}}}  + {\sigma _u},
\end{equation}
\noindent which can be calculated according to the difference between the total received reference signal power and the sum of the reported RSRPs~\cite{SECON2015}. This part is regarded as additive noise without fluctuation in the estimated SINR model.
\par
With the estimated parameters, i.e., ${{\hat p}_{u,b}}$, ${{p}'_{u,\sigma }}$ and ${p_{u,i}}$, $i\in{\bf{I'}}_u$, we can calculate the CDF and PDF of user CQI as follows,
\begin{equation}\label{eqSINRCDFMIAaprox}
 {F_{\Phi {'_u}}}\left( \phi  \right) = 1 - \exp \left( { - \frac{{{{p}'_{u,\sigma }}}}{{{{\hat p}_{u,b}}}}\phi } \right)\prod\limits_{i \in {\bf{I}}{'_u}} {{{\left( {\frac{{{p_{u,i}}}}{{{{\hat p}_{u,b}}}}\phi  + 1} \right)}^{ - 1}}},
   \end{equation}
\begin{equation}\label{eqSINRPDFMIAaprox}
 {f_{\Phi {'_u}}}\left( \phi  \right) = \left( {\frac{{{{p}'_{u,\sigma }}}}{{{{\hat p}_{u,b}}}} + \sum\limits_{i \in {\bf{I}}{'_u}} {\frac{{{p_{u,i}}}}{{{p_{u,i}}\phi  + {{\hat p}_{u,b}}}}} } \right)\left[ {1 - {F_{\Phi {'_u}}}\left( \phi  \right)} \right].
\end{equation}
\par
When all the IRSRPs are accessible and reported, the estimated SINR model is identical with that in~\eqref{eqSINRCDFMIA}, i.e.,
\begin{equation}\label{eqCDFrelation1}
{F_{{{\Phi '}_u}}}\left( \phi  \right) = {F_{{{\Phi}_u}}}\left( \phi  \right), \quad {{\bf{I}}^\prime }_u = {{\bf{I}}_u}.
\end{equation}
\noindent However, in reality, many weak interference signals cannot be identified by user terminals. Moreover, for the sake of less signalling overhead, $I_{max}$ is a limited number no larger than 8 according to~\cite{3GPP36331}. Therefore, we have the relationship as follows,
\begin{equation}\label{eqCDFrelation2}
{F_{{{\Phi '}_u}}}\left( \phi  \right) > {F_{{{\Phi}_u}}}\left( \phi  \right), \quad {{\bf{I}}^\prime }_u \subset {{\bf{I}}_u}.
\end{equation}
\noindent This means when partial IRSRPs are obtainable, the estimated CDF of user CQI in~\eqref{eqSINRCDFMIAaprox} is larger than the actual one, i.e., the estimated CQI is smaller. The proof of~\eqref{eqCDFrelation2} is presented in Appendix~C.
\par
Substituting~\eqref{eqSINRCDFMIAaprox} and~\eqref{eqSINRPDFMIAaprox} into~\eqref{eqwapproxi2}, we have the equations of the estimated user data rate, denoted as ${\overline r}'_u$. The closed-form solution of ${\overline r}'_u$ is intractable, we can nevertheless calculate the results by numerical methods~\cite{JIEA1992}. Thus, the overall throughput is calculated as
 \begin{equation}\label{}
{\overline r'_\Sigma } = \sum\limits_{u \in {\bf{U}}} {{{\overline r'}_u}}.
\end{equation}
 \noindent Although this analytical performance is the upper bound for PFS, it can be used as an estimation of user data rate and overall throughout in practical NOMA systems. In the following section, its estimation accuracy is evaluated by comparison with the simulation results.
\par
\section{Simulations and Numerical Results}
\begin{table}[t]
\centering
\caption{Simulation Parameters}\label{TableParameters}
\begin{tabular}{|c|c|}
 \hline {\textbf{Parameter}} & \textbf{Value} \\
 \hline Inter site distance & 500 m \\
 \hline Minimum link distance & 35 m \\
 \hline Bandwidth & 10 MHz @ 2.0 GHz\\
 \hline BS transmit power ($p_T$) & 46 dBm\\
 \hline BS transmit antenna gain & 15 dBi \\
 \hline Path loss & 128.1+37.6log($d$[km])\\
 \hline Standard deviation of shadow fading & 8 dB\\
 \hline Fast fading & Rayleigh model \\
 \hline Noise power density& -174 dBm/Hz \\
 \hline Noise figure & 5 dB \\
 \hline Frame duration & 10 $ms$ \\
 \hline Averaging window size ($\tau$)& 1000 \\
 \hline
\end{tabular}
\end{table}
In this section, the throughput performance of PFS in ideal and practical NOMA systems is evaluated by system-level simulations in Matlab. In practical NOMA systems with $S_{max} = 2$ and~$3$, the optimal multiplexed users and their assigned power are obtained by the tree searching-based user set selection scheme proposed in~\cite{ICC2016LF}. Then, the analytical performance of the upper bound is calculated and compared to the simulation results. The simulation parameters are configured according to~\cite{3GPP36839} and listed in Table~\ref{TableParameters}. A downlink cellular network with 37 cells is deployed in a hexagonal grid pattern. To avoid the edge effect, only the performance of the central cell is computed and other 36 BSs act as interferers. User terminals are uniformly randomly distributed in the cell. The long-term averaged user rates are initialized randomly, and we compute the statistic performance over 10,000 frames after 2,000 initial frames. The number of reported IRSRPs, i.e., $I_{max}$, is set to 8~\cite{3GPP36331}.
\par
\subsection{Throughput Performance of PFS}
\begin{figure}[t]
\centering
\subfigure[Overall throughput]{
\label{TPOverall}
\includegraphics[width=3.1in]{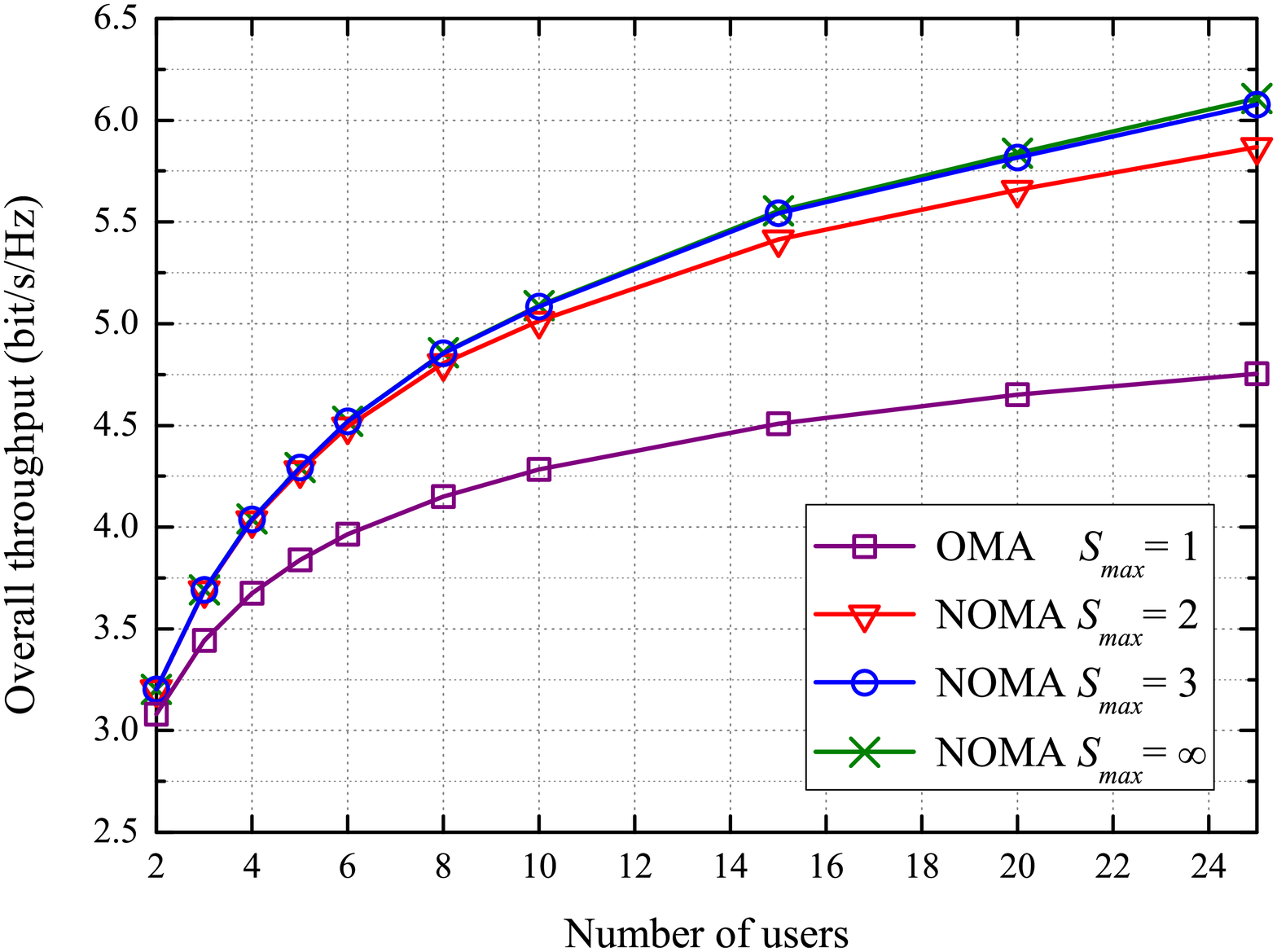}}
\subfigure[Cell-edge throughput]{
\label{TPCelledge}
\includegraphics[width=3.1in]{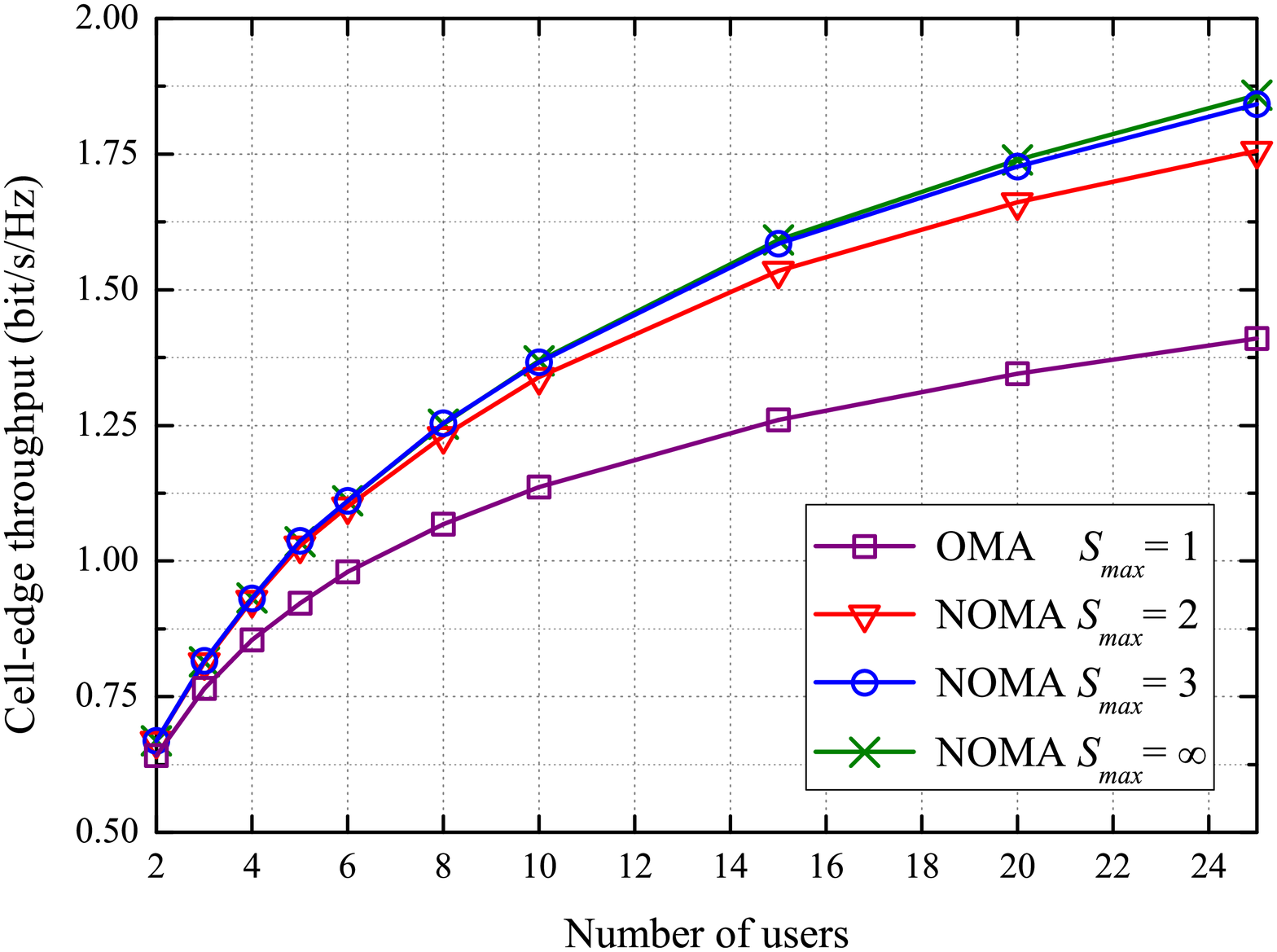}}
\caption{Simulation results of the throughput performance in NOMA and OMA systems.}
\label{FigTPperformance}
\end{figure}
The throughput performance of NOMA ($S_{max}>1$) and OMA ($S_{max} =1$) systems is presented in Fig.~\ref{FigTPperformance}. The cell-edge throughput shown in Fig.~\ref{TPCelledge} is defined as the mean rate of the lowest 5\% users. Compared to OMA, the NOMA system significantly improves the performance in terms of both spectrum efficiency and user fairness. The performance increases while there are more users in the cell, owing to the user diversity gain brought by PFS. Particularly, we have the ideal NOMA system with $S_{max} = \infty$. Consistent with our analysis in Section~III, it results in the highest throughput. The performance of PFS in practical NOMA ($S_{max}=2,\,3$) is extremely close to the upper bound, especially when there are fewer users. Due to this fact, it is feasible to use the analytical performance of the upper bound for data rate estimation in the practical system with a small $S_{max}$. In the following part, we investigate the relative deviation of the data rate estimation with various influence factors.
\par
\subsection{Number of Associated Users}
\begin{figure}[t]
\centering
\subfigure[Estimated overall throughput]{
\label{errorTPusernumber}
\includegraphics[width=3.1in]{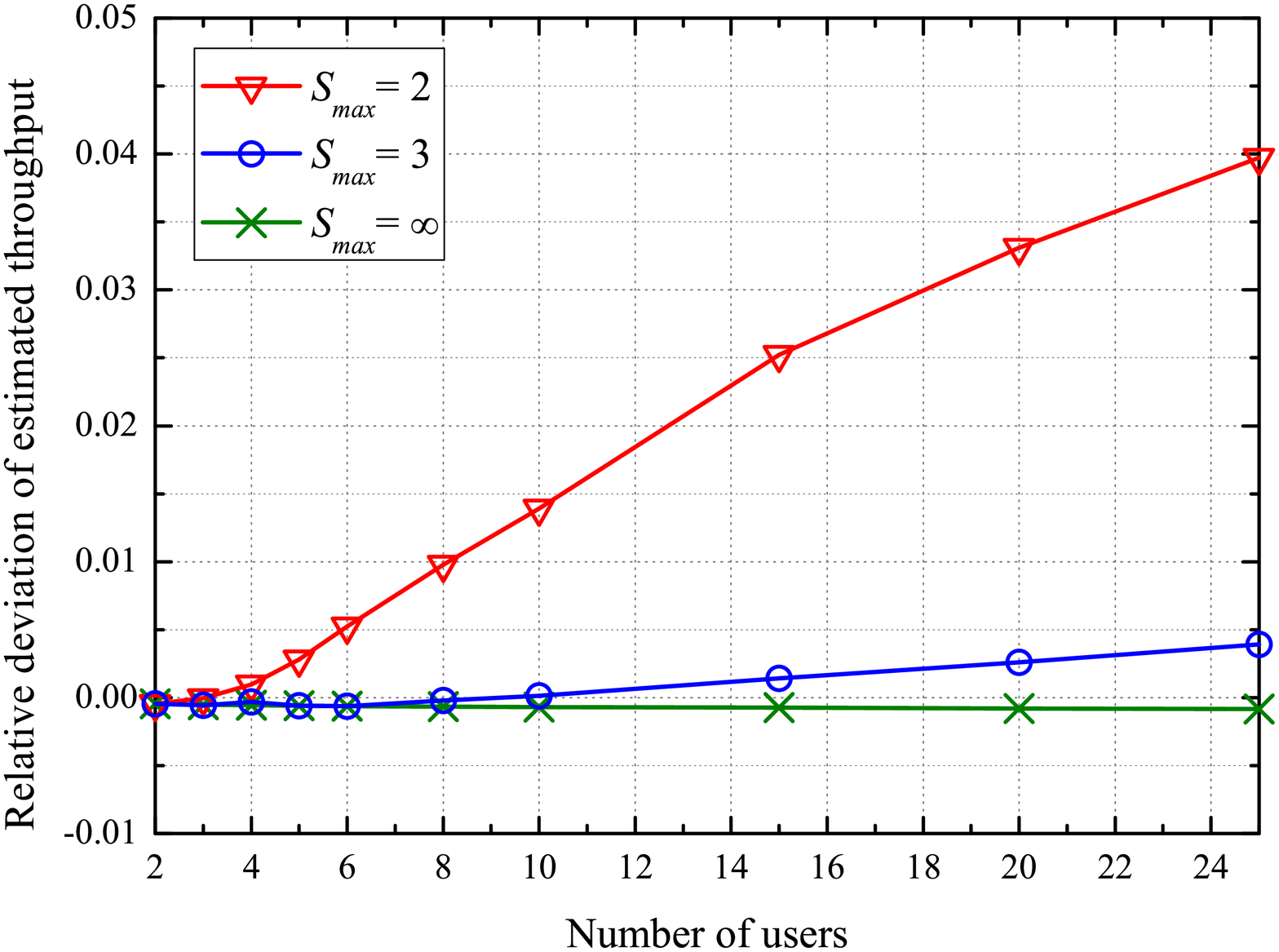}}
\subfigure[Estimated data rate per user]{
\label{errorCDFusernumber}
\includegraphics[width=3.1in]{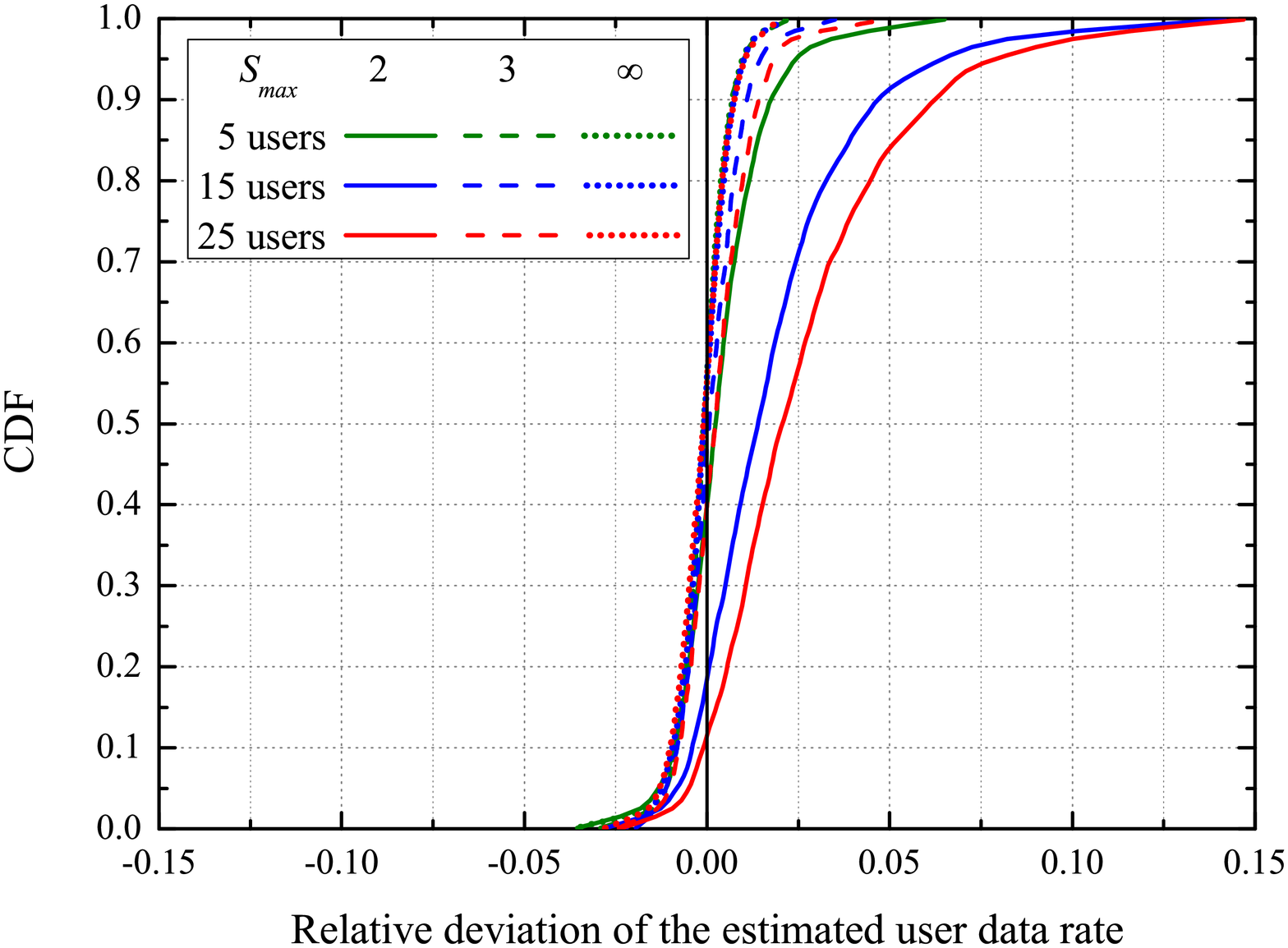}}
\caption{Relative deviation of the data rate estimation with different numbers of users.}
\label{Figerrorusernumber}
\end{figure}
In Fig.~\ref{errorTPusernumber}, we present the relative deviation of the estimated overall throughput, which is calculated as $\left( {{{\overline r'}_\Sigma } - {{\overline r}_\Sigma }} \right)/{\overline r_\Sigma }$. The analytical results are slightly lower than the simulation ones while $S_{max} = \infty$ due to the bias effect of the analytical SINR model with partial CQI information, which is explained in Section~\ref{SINRmodel}. The ideal NOMA system has the advantage that an arbitrary number of users can be multiplexed simultaneously. Therefore, its user diversity gain is higher than practical NOMA with a small $S_{max}$, especially when there are more users who have likely diverse CQIs. Nevertheless, the throughput estimation is very accurate even in the scenario with 25 users, where the relative deviation is lower than 0.005 for 3-user NOMA and is no more than 0.04 for 2-user NOMA.
\par
We calculate the relative deviation of the estimated data rate per user, i.e., $\left( {{{\overline r'}_u} - {{\overline r}_u}} \right)/{\overline r_u}$, and present its CDF in Fig.~\ref{errorCDFusernumber}. The estimated user date rates have extremely low deviations in the case that $S_{max} = \infty$, verifying our analytical result of the upper bound. The estimation results are very accurate when $S_{max} = 3$. This attributes to the smaller performance gap between the ideal and practical NOMA systems with $S_{max} = 3$, as shown in Fig.~\ref{FigTPperformance}. However, even when $S_{max}=2$, the estimation accuracy is very favorable for performance prediction. For instance, when there are 25 users, more than 97.5\% statistical relative deviations are within the range of~$\pm 0.10$.
\par
\subsection{Partial Channel Information}
\begin{figure}[t]
\centering
\subfigure[Estimated overall throughput]{
\label{errorTPReport}
\includegraphics[width=3.1in]{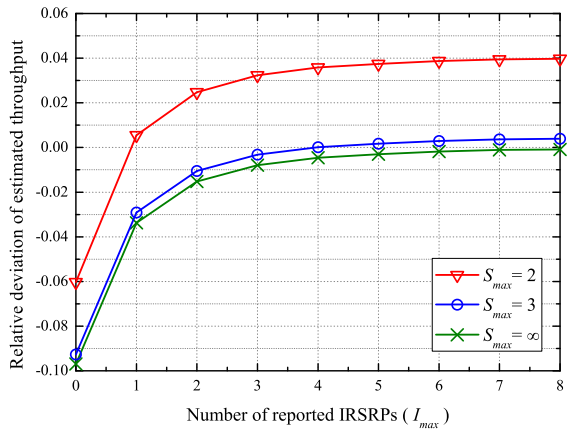}}
\subfigure[Estimated data rate per user]{
\label{errorCDFReport}
\includegraphics[width=3.1in]{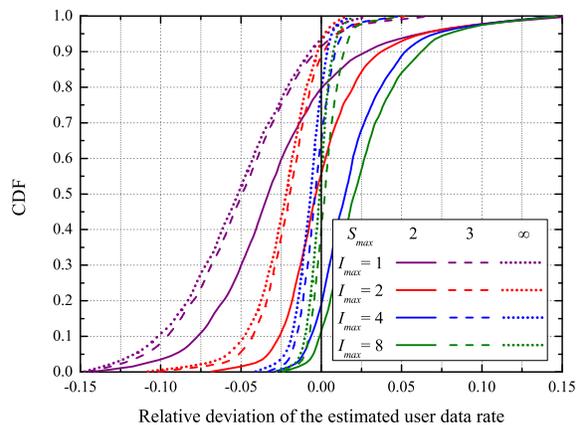}}
\caption{Relative deviation of the data rate estimation with different numbers of reported IRSRPs (25 users).}
\label{FigerrorReport}
\end{figure}
We investigate the influence of partial channel information on the estimation performance. The relative deviation of the estimated overall throughput with different numbers of reported IRSRPs is presented in Fig.~\ref{errorTPReport}. With fewer IRSRPs reported per user, the negative bias of the estimated SINR shown in~\eqref{eqCDFrelation2} becomes more evident. Thus, the estimated throughput is lower than the actual result when $I_{max}$ is small. The relative deviations are reduced within the range of $\pm 0.04 $ when $I_{max}\ge 1$, and barely change when $I_{max}>3$, indicating that the estimation of SINR distribution is accurate enough.
\par
We calculate the relative deviations of the estimated user data rates and their CDFs with different numbers of reported IRSRPs, as shown in Fig.~\ref{errorCDFReport}. When $I_{max} = 1$, the estimated user SINR distributions are inaccurate due to the lack of channel information. Thus, the relative deviations have large negative bias and wide extension. This inaccuracy drawback is remedied by increasing $I_{max}$ and becomes negligible in comparison with other influence factors (e.g., the number of users and $S_{max}$) when $I_{max}$ is larger than 4. As shown in Fig.~\ref{errorCDFReport}, the difference between the results with $I_{max} = 4$ and $8$ is less than 0.01. Therefore, it is reasonable to set $I_{max}$ to a smaller number than $8$ so that fewer IRSRPs are reported for the sake of reduction in signalling overhead.
\par
\subsection{Imperfect Channel Measurement}
\begin{figure}[t]
\centering
\subfigure[Estimated overall throughput]{
\label{errorTPCSI}
\includegraphics[width=3.1in]{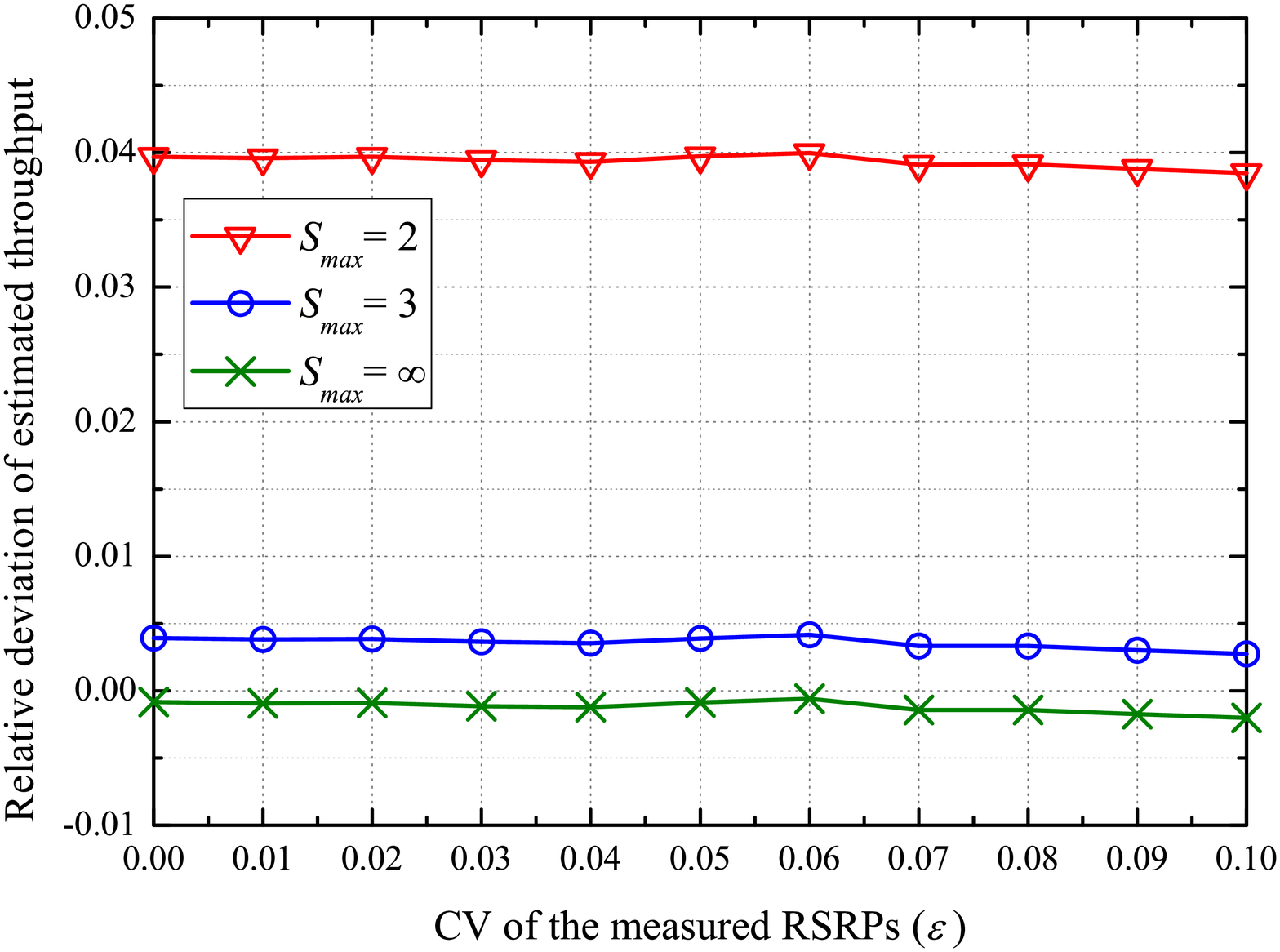}}
\subfigure[Estimated data rate per user]{
\label{errorCDFCSI}
\includegraphics[width=3.1in]{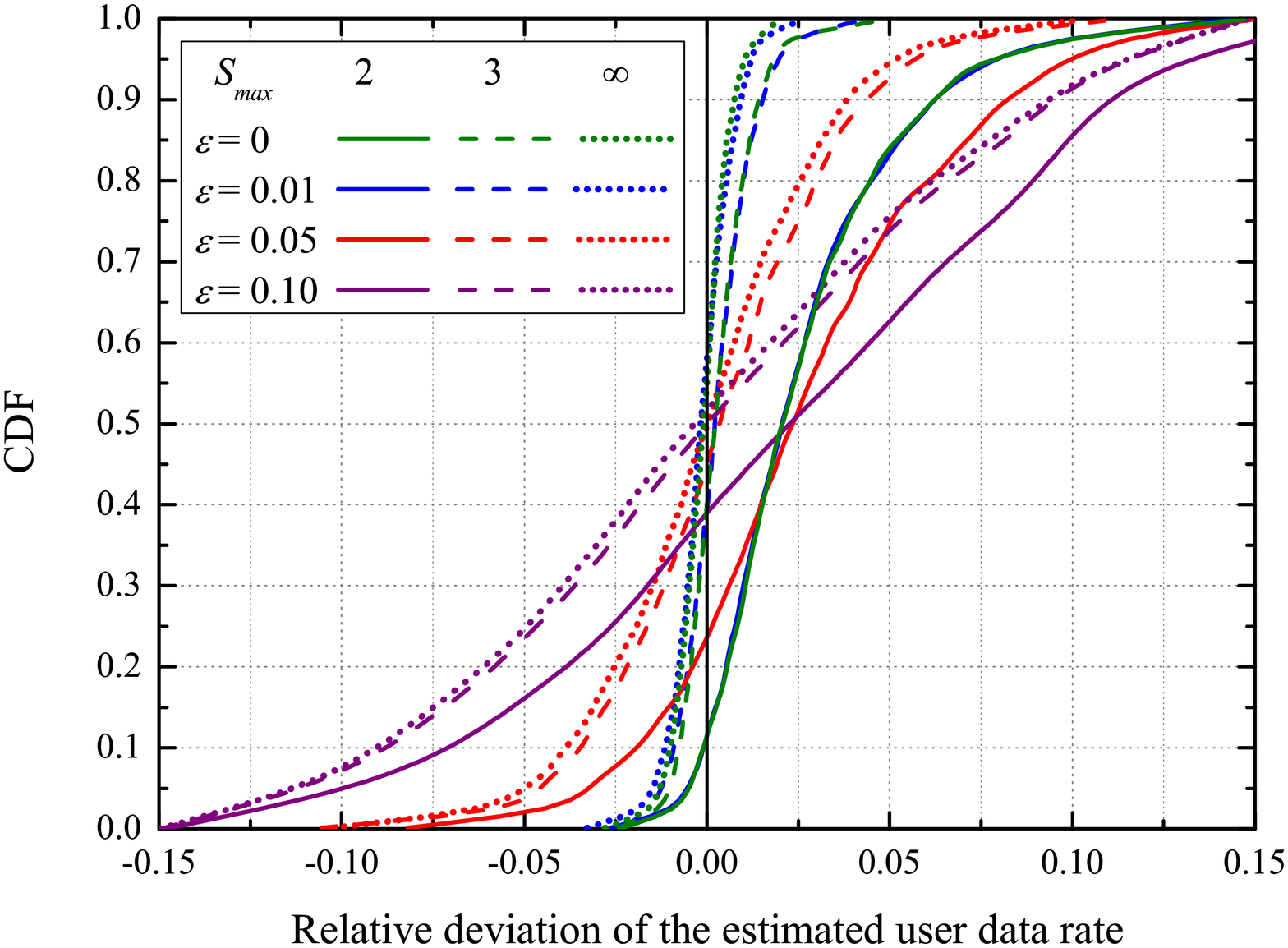}}
\caption{Relative deviation of the data rate estimation with different CVs of the measured RSRPs (25 users, $I_{max} = 8$).}
\label{FigerrorCSI}
\end{figure}
We consider the estimation deviation caused by the imperfect measurements of RSRPs. Each reported RSRP is the mean value of multiple samples of the received reference signal power. Under the Rayleigh fading channel, the reported RSRP follows the Erlang distribution. We denote its coefficient of variation (CV) as $\varepsilon>0$ for imperfect measurement. In Fig.~\ref{FigerrorCSI}, we present the relative deviation of the data rate estimation with different CVs of the measured RSRPs. The estimated overall throughput is insensitive to the inaccuracy of the measured RSRPs. However, as shown in Fig.~\ref{errorCDFCSI}, the CDFs of relative deviation per user indicate that the imperfect measurements enlarge the deviation range of the estimated user data rates. When $\varepsilon = 0.10$, only $81.2\%$ and $85.0\%$ statistical relative deviations are within the range of~$\pm 0.10$ in the 2-user and 3-user NOMA systems, respectively. As $\varepsilon$ is reduced by half, these proportions rise to $95.2\%$ and $99.2\%$. Thus, in order to improve the data rate estimation performance, high-precision channel measurement is always desired.
\par
\section{Conclusions}
In this paper, we analyzed the throughput performance of PFS for downlink NOMA systems. We derived a closed-form solution of the optimal PA for PFS based on the assumption of an arbitrary number of multiplexed users. Based on this optimal solution, we designed a low-complexity algorithm for joint power allocation and user set selection. The throughput performance of this optimal PA solution is analyzed based on the stochastic SINR model and acts as the upper bound for PFS. The simulation results verified our analysis on the upper bound performance. The analytical results are very close to the actual performance in the 2-user and 3-user NOMA systems. The relative deviation of the estimated data rate keeps at a significantly low level even when user amount increases to 25. Therefore, it is very feasible and accurate to use our analytical results for data rate estimation in practical NOMA. Moreover, we investigated the impact of partial and imperfect channel information on the estimation performance. With more than 4 reported IRSRPs, the negative bias effect of the SINR model can be relieved. Hence, only partial channel information is necessary for ensuring the estimation accuracy so that the signalling overhead can be reduced considerably. The imperfect RSRP measurement has negligible influence on the estimated overall throughput but leads to a larger deviation of the estimated data rate per user. Thus, it is necessary to enhance the channel measurement accuracy in order to improve the estimation performance.
\par
\numberwithin{equation}{section}
\appendices
\section{Proof of Theorem 1}
\begin{Proof}
Assuming ${\Phi _u} > {\Phi _v},u,v \in {\bf{U}}$, and letting ${\pi _u}\left( x \right) = {\pi _v}\left( x \right)$, we have
\begin{equation}\label{eqAPtheta}
x = {\theta _{u,v}} = \frac{{{R_u}\Phi _u^{ - 1} - {R_v}\Phi _v^{ - 1}}}{{{R_v} - {R_u}}}.
\end{equation}
\begin{enumerate}
\item \emph{Case 1}: If $0 < {\theta _{u,v}} < 1$, equally, we have
\begin{equation}
\left\{ {\begin{array}{*{20}{c}}
   {{\Phi _u} > {\Phi _v},} \notag \\
   {0 < \frac{{{R_u}\Phi _u^{ - 1} - {R_v}\Phi _v^{ - 1}}}{{{R_v} - {R_u}}} < 1} . \notag\\
\end{array}} \right.
\end{equation}
\begin{equation}\label{eqAPthetarange}
\Leftrightarrow 0 < \frac{{{\Phi _v}}}{{{\Phi _u}}} < \frac{{{R_v}}}{{{R_u}}} < \frac{{1 + \Phi _u^{ - 1}}}{{1 + \Phi _v^{ - 1}}} < 1.
\end{equation}
Under this condition,
\begin{itemize}
  \item when $x \in \left( {0,{\theta _{u,v}}} \right)$, it holds that
  \begin{align}\label{}
 &{\pi _u}\left( x \right) - {\pi _v}\left( x \right) \notag\\
&  = \frac{{{R_v}\Phi _v^{ - 1} - {R_u}\Phi _u^{ - 1} - x\left( {{R_u} - {R_v}} \right)}}{{{R_u}{R_v}\left( {\Phi _u^{ - 1} + x} \right)\left( {\Phi _v^{ - 1} + x} \right)}} \notag\\
 &\mathop  > \limits^{\eqref{eqAPthetarange}} \frac{{{R_v}\Phi _v^{ - 1} - {R_u}\Phi _u^{ - 1} - {\theta _{u,v}}\left( {{R_u} - {R_v}} \right)}}{{{R_u}{R_v}\left( {\Phi _u^{ - 1} + x} \right)\left( {\Phi _v^{ - 1} + x} \right)}} \notag\\
  & \mathop  = \limits^{\eqref{eqAPtheta}} 0 ;\notag
  \end{align}
  \item when $x \in \left( {{\theta _{u,v}}},1 \right)$, on the contrary, it holds that
\end{itemize}
\begin{equation}\label{}
{\pi _u}\left( x \right) - {\pi _v}\left( x \right)<0. \notag
\end{equation}
\item \emph{Case 2}: If user $u$, $v$ have relationship as
    \begin{equation}
    \frac{{{R_v}}}{{{R_u}}} \le \frac{{{\Phi _v}}}{{{\Phi _u}}}, \notag
  \end{equation}
  then we have
\begin{equation}
\frac{{{\pi _u}\left( x \right)}}{{{\pi _v}\left( x \right)}} = \frac{{{\Phi _u}}}{{{\Phi _v}}}\frac{{{R_v}}}{{{R_u}}}\frac{{\left(1 + x{\Phi _v}\right)}}{{\left(1 + x{\Phi _u}\right)}} < 1. \notag
\end{equation}
\item \emph{Case 3}: If user $u$, $v$ have relationship as
\begin{equation}
\frac{{{R_v}}}{{{R_u}}} \ge \frac{{1 + \Phi _u^{ - 1}}}{{1 + \Phi _v^{ - 1}}}, \notag
\end{equation}
then we have
\begin{align}
 &\frac{1}{{{\pi _u}\left( x \right)}} - \frac{1}{{{\pi _v}\left( x \right)}}\notag \\
  = & \left( {1 + \Phi _u^{ - 1}} \right){R_u} - \left( {1 + \Phi _v^{ - 1}} \right){R_v} + \left( {1 - x} \right)\left( {{R_v} - {R_u}} \right) \notag\\
  \le &\left( {1 - x} \right)\left( {\frac{{1 + \Phi _u^{ - 1}}}{{1 + \Phi _v^{ - 1}}} - 1} \right){R_u}   \notag\\
  <& 0 .\notag
\end{align}
\end{enumerate}
So far, the three cases in Theorem 1 are proved completely.
\end{Proof}
\section{}
\begin{Proof}
According to the definition in \eqref{eqPFR}, the expectation of the long-term averaged data rate is expressed as
\begin{equation}\label{eqAPeR}
\mathbb{E}\left[ {{R_u}\left( {t + 1} \right)} \right] = \left( {1 - \frac{1}{\tau }} \right)\mathbb{E}\left[ {{R_u}\left( t \right)} \right] + \frac{1}{\tau }\mathbb{E}\left[ {{r_u}\left( t \right)} \right].
\end{equation}
\par
Assuming ergodicity for $R_u\left( t \right)$ for stable PFS, it holds that
\begin{equation}\label{eqAPRR}
\mathbb{E}\left[ {{R_u}\left( {t + 1} \right)} \right] = \mathbb{E}\left[ {{R_u}\left( t \right)} \right].
\end{equation}
\par
Substituting~\eqref{eqAPRR} into~\eqref{eqAPeR}, we have
\begin{equation}\label{eqAPB3}
\mathbb{E}\left[ {{R_u}\left( t \right)} \right] = \mathbb{E}\left[ {{r_u}\left( t \right)} \right] = {{\overline r}_u}
\end{equation}
\par
When the averaging window size $\tau \gg 1$, we have the approximation as
\begin{equation}\label{eqAPB4}
\frac{1}{\tau ^{n}} \approx 0,\quad n \ge 2.
\end{equation}
\par
Considering again the ergodicity of $R_u\left( t \right)$, we assume that for a certain frame $t$ there exists
\begin{equation}\label{eqAPB5}
{R_u}\left( {{t} + N} \right) = {R_u}\left( {{t}} \right),
\end{equation}
\noindent which means that the status ${R_u}\left( {t} \right)$ repeats after a long enough period, i.e, $N$ frames after frame $t$. Then, we can deduce ${R_u}\left( {t} \right)$ as follows,
\begin{align}\label{eqAPB6}
 &{R_u}\left( {{t} + N} \right) \notag\\
 &= {\left( {1 - \frac{1}{\tau }} \right)^N}{R_u}\left( {{t}} \right) + \sum\limits_{n = 0}^{N - 1} {\frac{1}{\tau }{{\left( {1 - \frac{1}{\tau }} \right)}^{N-n-1}}{r_u}\left( {{t} + n} \right)} \notag\\
 &\mathop  \approx \limits^{\eqref{eqAPB4}} \left( {1 - \frac{N}{\tau }} \right){R_u}\left( {{t}} \right) + \sum\limits_{n = 0}^{N - 1} {\frac{1}{\tau }{r_u}\left( {{t} + n} \right)} ,
 \end{align}
 \begin{equation}\label{eqAPB7}
 \mathop  \Rightarrow \limits^{\eqref{eqAPB5}} {R_u}\left( {{t}} \right) \approx \frac{1}{N}\sum\limits_{n = 0}^{N - 1} {{r_u}\left( {{t} + n} \right)}  \approx {{\overline r}_u}.
\end{equation}
\par
Combining~\eqref{eqAPB3} and~\eqref{eqAPB7}, it is proved that
\begin{equation}\label{eqAPB8}
{R_u}\left( t \right) \approx \mathbb{E}\left[ {{R_u}\left( t \right)} \right] = {{\overline r}_u} ,\quad \forall t.
\end{equation}
\end{Proof}
\section{}
\begin{Proof}
Considering the property that
\begin{equation}\label{eqC1}
e^x > 1 + x,\quad x > 0,
\end{equation}
\noindent the relationship in~\eqref{eqCDFrelation2} can be proved as follows,
\begin{align}
 {F_{\Phi {'_u}}}\left( \phi  \right) &= 1 - \exp \left( { -\phi \frac{{{{p}'_{u,\sigma }}}}{{{{\hat p}_{u,b}}}} } \right)\prod\limits_{i \in {\bf{I}}{'_u}} {{{\left( {\frac{{{p_{u,i}}}}{{{{\hat p}_{u,b}}}}\phi  + 1} \right)}^{ - 1}}} \notag \\
&= 1 - \exp \left( { - \phi \frac{{{\sigma _u}}}{{{{\hat p}_{u,b}}}}} \right)\frac{{{\prod \limits_{i \in {{{\bf{I'}}}_u}}}{{\left( {{p_{u,i}}\hat p_{u,b}^{ - 1}\phi  + 1} \right)}^{ - 1}}}}{{{\prod \limits_{i \in \left( {{{\bf{I}}_u} - {{{\bf{I'}}}_u}} \right)}}{\exp \left( {{p_{u,i}}\hat p_{u,b}^{ - 1}\phi } \right)}}} \notag \\
& \mathop > \limits^{\eqref{eqC1}} 1 - \exp \left( { - \phi \frac{{{\sigma _u}}}{{{{\hat p}_{u,b}}}}} \right)\frac{{{\prod \limits_{i \in {{{\bf{I'}}}_u}}}{{{\left( {{p_{u,i}}\hat p_{u,b}^{ - 1}\phi  + 1} \right)}}^{ - 1}}}}{{{\prod \limits_{i \in \left( {{{\bf{I}}_u} - {{{\bf{I'}}}_u}} \right)}}{\left( {{p_{u,i}}\hat p_{u,b}^{ - 1}\phi  + 1} \right)}}}  \notag \\
  &= {F_{{\Phi _{u}}}}\left( \phi  \right). \notag
\end{align}
\end{Proof}

\section*{Acknowledgment}
The authors would like to thank Prof. Petri~M\"{a}h\"{o}nen for the fruitful discussion and his constructive feedback on this work.
\par

\end{document}